\documentclass{emulateapj}
\usepackage[utf8]{inputenc} 
\usepackage{amsmath}
\usepackage{amssymb}
\usepackage{graphicx}
\usepackage{grffile}
\usepackage[usenames,dvipsnames]{color}
\usepackage{bm}
\usepackage{natbib}
\usepackage{hyperref}
\usepackage{tabularx}
\usepackage{xspace} % smart spaces

\usepackage[english]{babel}

\newcommand{\dd}{\mathrm{d}}
\newcommand{\vek}[1]{\mathbf{#1}}

\newcommand{\pow}[1]{\ifmmode{}^{#1}\else ${}^{#1}$\fi}
\newcommand{\HI}{{\text{H\MakeUppercase{\romannumeral 1}}}\xspace}

\newcommand{\Lya}{\ifmmode{\mathrm{Ly}\alpha}\else Ly$\alpha$\xspace\fi}

\newcommand{\cm}{\,\ifmmode{{\rm cm}}\else cm\fi}

\newcommand{\ergps}{\,{\rm erg}\,{\rm s}\ifmmode{}^{-1}\else ${}^{-1}$\fi}
\newcommand{\Mpch}{\,{\rm Mpc}\,\ifmmode h^{-1}\else $h^{-1}$\fi}
\newcommand{\snru}{\,\ifmmode{\mathrm{Myr}^{-1}}\else Myr${}^{-1}$\fi}
\newcommand{\kms}{\,\ifmmode{\mathrm{km}\,\mathrm{s}^{-1}}\else km\,s${}^{-1}$\fi}

\newcommand{\silcc}{\texttt{SILCC}\xspace}

\slugcomment{Draft from \today}

\begin{document}

\title{The imprint of cosmic ray driven outflows on Lyman-$\alpha$ spectra}

\author{Max Gronke$^{1}$}
\author{Philipp Girichidis$^{2}$}
\author{Thorsten Naab$^{3}$}
\author{Stefanie Walch$^{4}$}
\affil{$^{1}${Department of Physics, University of California, Santa Barbara, CA 93106, USA; }}
\affil{$^{2}${Leibniz-Institut f\"ur Astrophysik Potsdam, An der Sternwarte 16, D-14482 Potsdam, Germany; }}
\affil{$^{3}${Max-Planck-Institut f\"ur Astrophysik, Karl-Schwarzschild-Str. 1, D-85741 Garching, Germany; }}
\affil{$^{4}${I. Physikalisches Institut, Universita\"at zu K\"oln, Z\"ulpicher Str. 77, D-50937 K\"oln, Germany.}}

\email{maxbg@ucsb.edu}

\begin{abstract}
  Recent magneto-hydrodynamic simulations of the star-forming interstellar medium (ISM) with parsec scale resolution indicate that relativistic cosmic rays support the launching of galactic outflows on scales of a few kpc. If these fundamental constituents of the ISM are injected at the sites of supernova (SN) explosions, the outflows are smoother, colder, and denser than the highly structured, hot-phase driven outflows forming, e.g., by thermal SN energy injection alone. In this Letter we present computations of resonant Lyman-$\alpha$ (Ly$\alpha$) radiation transfer through snapshots of a suite of stratified disk simulations from the \texttt{SILCC} project. For a range of thermal, radiative, and kinetic feedback models only simulations including non-thermal cosmic rays produce Ly$\alpha$ spectra with enhanced red peaks and strong absorption at line center -- similar to observed systems. The absence of cosmic ray feedback leads to spectra incompatible with observations. We attribute this to the smoother neutral gas distribution of cosmic ray supported outflows within a few kpc from the disk midplane.
\end{abstract}


\section{Introduction}
\label{sec:intro}
Galactic outflows are commonly detected through absorption and emission line studies \citep{2005ARA&A..43..769V,Steidel2010ApJ...717..289S,Tumlinson2017}.
A particularly noteworthy emission line is Lyman-$\alpha$ (\Lya), which is due to the first transition of atomic hydrogen. \citet{Partridge1967} realized the potential use of this bright emission line to detect star-forming galaxies, and since then thousands of \Lya emitting galaxies have been found. However, its use in probing the gas kinematics is limited due to the resonant nature of \Lya, which leads to scattering processes and frequency shifts. Thus, the observed spectrum cannot be directly linked to the kinematics of the emitting regions as for optically thinner lines.

The resonant nature of \Lya is often seen as a major shortcoming but it holds an immense potential since the \Lya photons sample -- in their semi-random walk through the galaxy -- the density and kinematic structure of neutral hydrogen (\HI),
which are difficult to probe otherwise \citep{Neufeld1990,Eide2018}.

Comparisons of observed with synthetic \Lya spectra yield that the majority of spectral shapes are consistent with outflows, and are reproducible even with very simplified models \citep[e.g.,][]{Gronke2017a}. These functioning models have usually a homogeneous geometry such as a slab or a shell -- which is somewhat surprising, as observations \citep{2005ARA&A..43..769V}, analytic considerations \citep{McKee1977} as well as numerical simulations \citep[e.g.,][]{Walch2015,2018ApJ...853..173K} show that the interstellar medium (ISM) and its outflows have a multiphase structure.

The hot phase of this medium -- one major driver for outflows -- is mainly generated by supernovae (SN) explosions, which dump
large amounts of energy into their surroundings \citep[see, e.g.,][]{Naab2016}.
Recent ISM scale simulations presented in \citet{Girichidis2018} highlight that the inclusion of cosmic rays (CRs), in addition to SN explosions, change the multiphase structure of the outflows. They become smoother and overall denser transporting more $\sim 10^4\,$K gas away from the star-forming disk. Here, we investigate the effect of CR driven outflow features on \Lya observables. We, therefore, use mainly the simulations of \citet{Girichidis2018} (see \S\ref{sec:method}).

\begin{figure*}
  \centering
  \includegraphics[width=.95\textwidth]{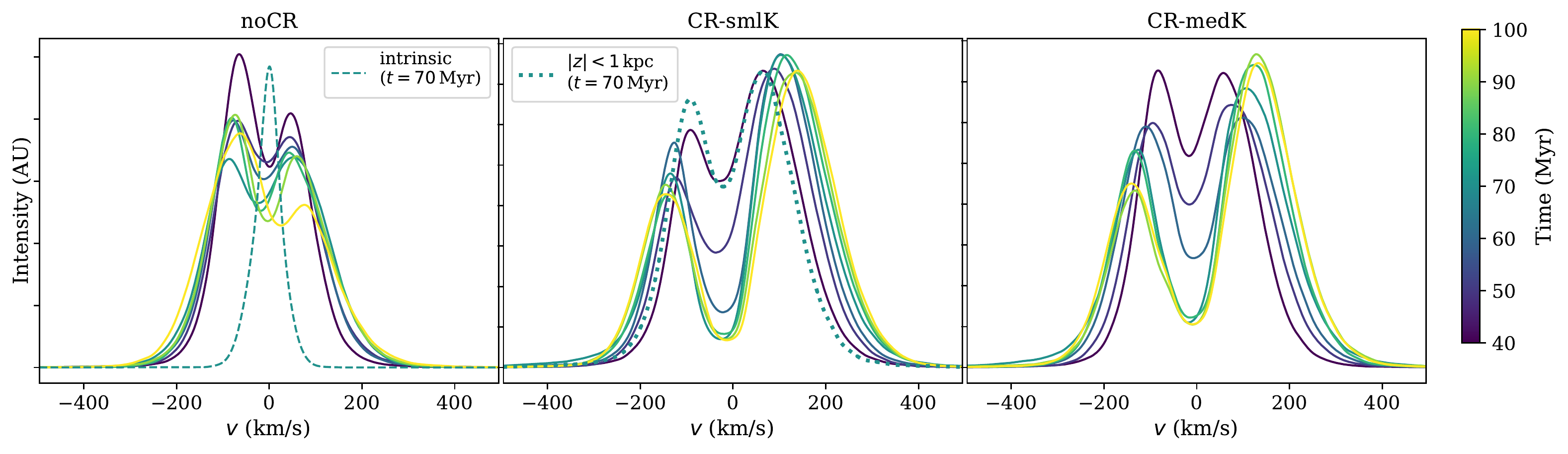}
  \vspace{-0.1cm}
  \caption{Emergent \Lya spectra of the simulations without CRs \textit{(left)} and with CR feedback with small and medium diffusion coefficient (\textit{center and right}). Shown are the spectra from $t=40\,$Myr to $t=100\,$Myr in steps of $10\,$Myr after the start of the simulation. Only simulations with CRs show the dominant red peak ($v>0$), and a strong depression at line center at times when the outflow is fully developed ($t \gtrsim 70\,$Myr). As examples, the \textit{dashed line} shows an intrinsic spectrum which is widened through radiative transfer effects, and the \textit{dotted line} in shows the spectrum emergent from the inner $1\,$kpc (both at $70\,$Myr), i.e., before the photons have to cross the CR driven outflow.}
  \label{fig:spectra}
\end{figure*}

\begin{figure*}
  \centering
\includegraphics[height=15cm]{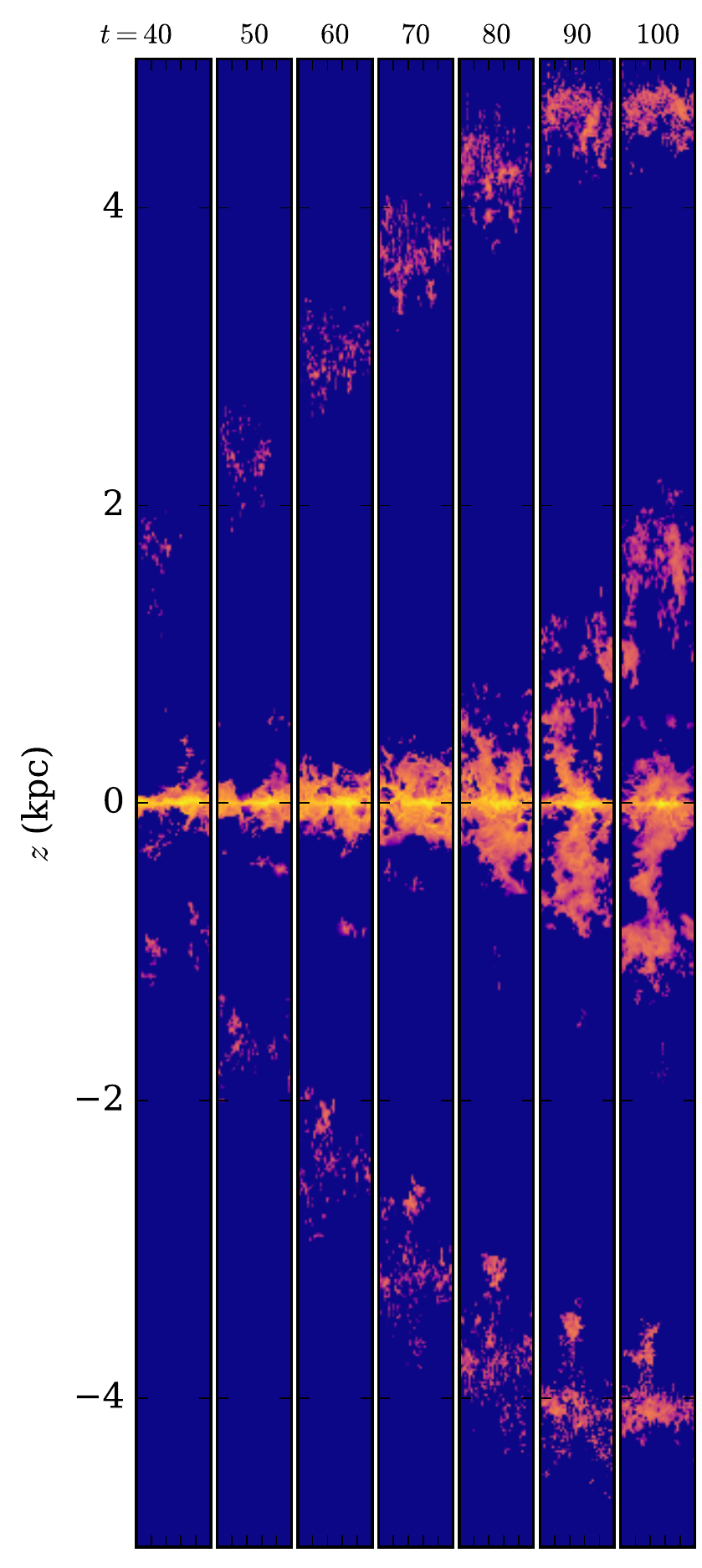}\qquad\includegraphics[height=15cm]{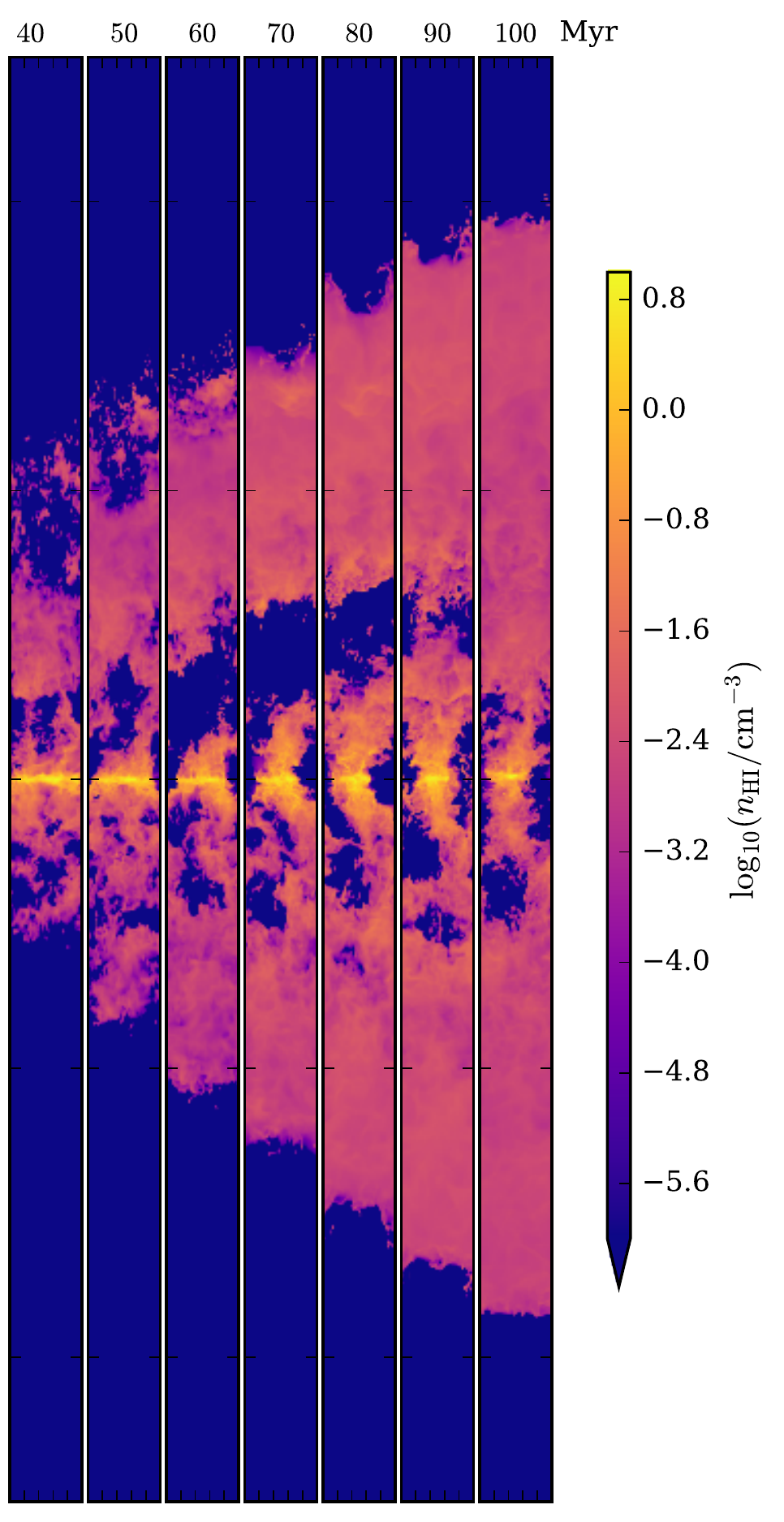}
  \caption{Projected $n_{\HI}$ distribution of the `noCR' (\textit{left}) and `CR-smlK'  simulations (\textit{right}) which shows a smoother and more volume filling \HI distribution for the latter. The position of last scattering follows roughly the $n_\HI$ distribution. Thus, the observable surface brightness distribution would be much more spread out in the `CR-smlK' case.}
  \label{fig:grids}
\end{figure*}

\section{Method}
\label{sec:method}

\subsection{The SILCC simulations}
\label{sec:silcc}
The simulation setups are described in detail in \citet{Walch2015} and \citet{Girichidis2016}. For the details of the CR implementation we refer to \citet{Girichidis2016a} and \citet{Girichidis2018}. In the following we only give a brief summary. The magneto-hydrodynamic (MHD) simulations are carried out using the \texttt{FLASH} code \citep{2000ApJS..131..273F,DubeyEtAl2009}\footnote{\url{http://flash.uchicago.edu}}. The representative volume of the SN-driven ISM is a stratified box with a size of $0.5\,\mathrm{kpc} \times 0.5\,\mathrm{kpc} \times\pm5\,\mathrm{kpc}$. The MHD equations are solved using the HLLR3/5 solver \citep{Waagan2011JCoPh.230.3331W}, which has been extended to include CRs as a relativistic fluid in the advection-diffusion approximation. Gravitational attraction includes the effects of self-gravity using a tree-based method by \citet{WuenschEtAl2018} as well as an external potential to account for the stellar component of the disk. The thermodynamical evolution is coupled to a chemical network based on \citet{GloverClark2012b} which follows the evolution of hydrogen in the form of $\mathrm{H_2},\ \mathrm{H,}\ \mathrm{or}\ \mathrm{H^+}$. As a dynamical driver we use SNe that explode at a constant rate. We consider clustering of the SNe based on observational constraints. For every SN we inject a thermal energy of $10^{51}\,$erg if the Sedov-Taylor radius is resolved with at least $4$ cells (which is the case for $\gtrsim 95\%$), otherwise we inject the terminal momentum \citep[see][]{2015MNRAS.449.1057G,2016MNRAS.460.2962H}. In the case of CRs, we additionally inject $10^{50}$ erg as CRs. We run the simulations for $100-150$\,Myr.

We mainly use the simulations `noCR', `CR-smlK', and `CR-medK' of \citet{Girichidis2018}, which feature no CR feedback, and two different CR diffusion coefficients. For comparison, we also use simulations with a more self-consistent cluster based star formation algorithm and the inclusion of stellar wind \citep{Gatto2017} and ionizing radiation \citep{Peters2016}.

\subsection{Radiative transfer}
\label{sec:rt}
We use the Monte-Carlo radiative transfer code \texttt{tlac} \citep{Gronke2014a} to follow the trajectories and frequencies of individual photon packages. The emergent spectrum is the distribution of the photons' frequencies escaping the simulation box \citep[e.g.,][]{MarkLectureNotes}.

Prior to the radiative transfer calculations, we convert the \silcc output to a uniform grid with cellsize $\sim 4\,$pc, corresponding to the spacing of the maximum refinement of the hydrodynamical simulation. We associate each cell with the following properties: 
\begin{itemize}
\item the neutral hydrogen number density $n_{\HI}$, which we take directly from the \silcc output,
\item the effective temperature $T$, which is the cell temperature from the hydrodynamical simulation, 
  
\item the dust opacity per cell which we calculate following \citet{Laursen2009} as $\kappa_d = \sigma_{\rm SMC}(n_\HI + a_{\rm ion} n_{\rm HII})$,
where $\sigma_{\Lya, {\rm SMC}}\approx 1.58\times 10^{-21}\,{\rm cm}^{2}$ is the \Lya dust cross section for the Small Magellanic Cloud \citep{Pei1992}, $n_{\mathrm{HII}}$ the $\mathrm{H}^+$ number density, and we use a dust-to-gas ratio in the ionized regions of $a_{\rm ion}=0.01$,
\item the gas bulk velocity $\vek{v}$ from the hydrodynamical simulations, and
  
\item the \Lya emissivity $\epsilon = n_{\rm HII} n_{\rm e} N_\alpha(T) \alpha_{\rm B}(T)$ where $N_\alpha(T)$ is the number of \Lya photons produced per recombination event for which we use the fit provided by \citet{Cantalupo2008}, and $\alpha_{\rm B}(T)$ is the `case-B' recombination coefficient as approximated by \citet{Hui1997}.
\end{itemize}
For each photon package we draw the starting position randomly proportional to a cell's emissivity\footnote{Within a given cell, we distribute the starting position uniformly.}. The initial frequency is drawn from a Voigt function with mean and width corresponding to the bulk velocity and approximately the thermal velocity of \HI, $v_{\mathrm{th}}$, of the starting cell, respectively. A random initial direction (isotropically distributed), and a `travelling optical depth', $\tau$, (from an exponential distribution with unity scale) are drawn. We compute the photon's travelling distance $d$ via
\begin{equation}
  \label{eq:lya_RT_dist}
  \tau = \int\limits_0^{d}\dd s \left(n_{\HI}\sigma_{\HI}(v,T) +  \kappa_{\mathrm{d}}\right),
\end{equation}
where  $\sigma_{\HI}(v,T)$
is the \Lya scattering cross section in units of velocity offset from line-center $v=(\nu_{\Lya}/\nu - 1)c$. Afterwards, the photon scatters off hydrogen with a probability $\sigma_\HI n_{\HI}$, or off dust (with a probability $A_{\mathrm{d}} \kappa_{\mathrm{d}}$ where $A_{\mathrm{d}}=0.32$ is the dust albedo), or is absorbed by dust.
After scattering, we draw the new photon direction from the corresponding phase function \citep[see][]{Gronke2014a}.

It is possible to omit the scatterings near line-center leading to essentially no displacement of the photon. We use a (conservative) dynamic core-skipping technique following \citet{Smith2014} with an upper limit $v_{\mathrm{crit., max}}\sim 10 v_{\mathrm{th}}$. We compared some selected spectra to the results obtained for simulations without speeding up the calculations, and found them to be indistinguishable.

\begin{figure*}
  \centering
  \includegraphics[width=.95\linewidth]{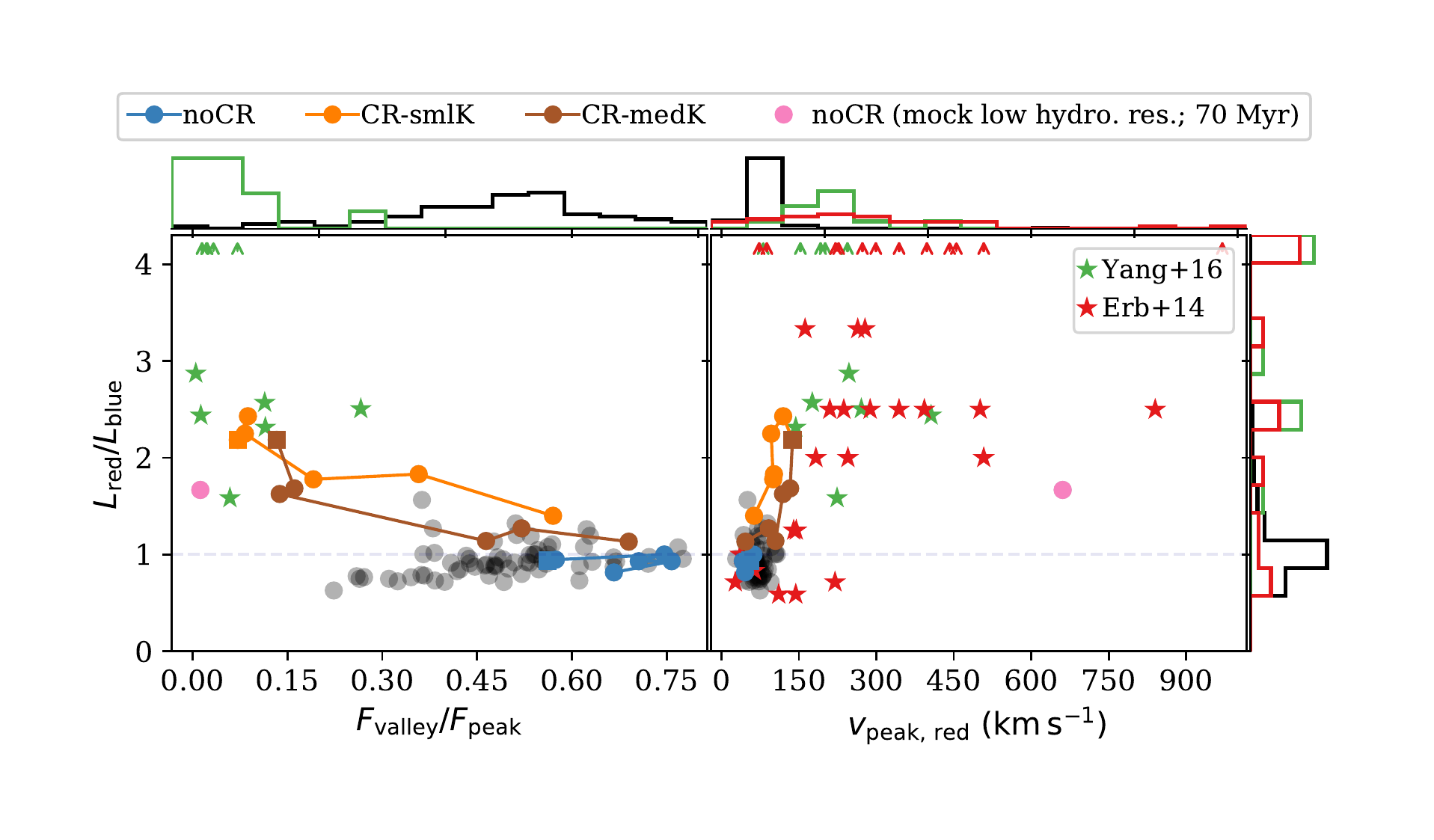}
  \vspace{-0.8cm}
  \caption{Comparison of the modelled \Lya spectra (circles) to observations (stars). We characterize the spectral shape with the asymmetry of the line ($y$-axis) versus the `deepness' of the draught (\textit{left panel}) and the red peak's position (\textit{right panel}). In each panel, we highlight our simulated spectra from $40\,$Myr to $100\,$Myr by color with the last data point marked with a square. The black circles are other \silcc simulations shown for comparison (see \S~\ref{sec:results} for details). Note the arrows on the top border of each panel showing the observations with $L_{\mathrm{red}}/L_{\mathrm{blue}}> 4$. In general, only simulations with CRs result in $L_{\mathrm{red}}/L_{\mathrm{blue}}$ and $F_{\mathrm{valley}}/F_{\mathrm{peak}}$ in agreement with observations. The red peak positions are low in all cases.}
  \vspace{.5cm}
  \label{fig:obs}
\end{figure*}

\section{Results}
\label{sec:results}
Fig.~\ref{fig:spectra} shows the computed \Lya spectra for three simulations introduced in \citet{Girichidis2018}.
In each panel, we show the spectra of the snapshots at $t=40,\,50\,\ldots\,100\,$Myr (color-coded) assembled with the photons escaping in the positive $z$-direction (the \Lya escape fractions are typically $\sim 30-80\,\%$). The escape directions are as for an optically thick slab, i.e., proportional to the cosine of the viewing angle.
Each spectrum consists of $10^5$ escaped photon packages and was smoothed with a Gaussian kernel (standard deviation $\sigma=10\kms$).
The left panel shows the results of the `noCR' simulation (only thermal feedback; where we also included as an example an intrinsic spectrum as dashed line) whereas the other two panels show the results of two simulations including CRs. While the spectra of the former simulations stay similar in shape with a fairly symmetric or even dominant blue ($v < 0$) side, the spectra of the CR simulations evolve over time with an increasingly dominant red ($v>0$) side and a deeper ``draught'' between the two peaks. The spectra of the `downward' side (photons escaping in a negative $z$-direction; not shown) look very similar.  The `noCR' spectra look quantitatively similar to spectra from similar simulations including self-gravity \citep{Walch2015,Girichidis2016}, additional feedback from stellar winds \citep{Gatto2017}, and ionizing radiation \citep{Peters2016}.

This difference in the respective \Lya spectra caused by the CR feedback can be understood when analyzing the morphology of the neutral gas. In Fig.~\ref{fig:grids} we show the two-dimensional projection of the neutral hydrogen in our simulation boxes with (right) and without CRs (left). Clearly visible in both cases is the dense, central disk extending over $z\lesssim 0.2\,$kpc. There is additional  neutral gas at larger heights which has been expelled from the inner disk. This gas is dispersed and clumpy in the purely thermal feedback case, and in a more homogeneous `shield' in the simulation including CRs.
As discussed in detail in \citet{Girichidis2018} the CRs provide a much smoother outflow leading, for instance, to clumping factors more than one order of magnitude lower than in the `noCR' case (\citealp{Girichidis2018}, figure 11). In the central panel of Fig.~\ref{fig:spectra}, we include an example spectrum where we cut out the gas at $|z|>1\,$kpc. In that case, when the photons do not have to cross through the CR driven outflow, the spectrum resembles closely the `noCR' case.

To compare our findings to observations, we extract the following characteristics from the computed spectra: \textit{(i)} the ratio of the integrated redward over the integrated blueward flux ($L_{\mathrm{red}}/L_{\mathrm{blue}}$) -- which is a measure of the asymmetry of the line shape; \textit{(ii)} the flux at `draught' between the peaks ($F_{\mathrm{valley}}$) in comparison to the flux at the maximum point ($F_{\mathrm{peak}}$), which, in practice, quantifies the `deepness' of the draught of the double peaked spectra; and \textit{(iii)} the position of the red emission peak ($v_{\mathrm{red}}$). We show the evolution of these measures in Fig.~\ref{fig:obs} from $t=40$ to $100\,$Myr (in intervals of $10\,$Myr) with the last data point marked with a square.
We compare this to observations of nearby, star-forming galaxies (\citealp[the `Green Peas'; data taken from][]{Yang2015}, also \citealp{Henry2015}), and to $z\sim 2-3$ Lyman-$\alpha$ selected galaxies \citep{Erb2014}.
In addition, we also display other simulation results presented in \citet{Walch2015}, \citet{Gatto2017}, and \citet{Peters2016} (shown as black circles in Fig.~\ref{fig:obs}). These hydrodynamical simulations include supernovae feedback with random SN positions and cover SN rates from $5$ to $45\,\mathrm{Myr}^{-1}$ with none including CRs (but partially include stellar wind and UV feedback). We show them only to support the case that the spectral shape we obtain for the `noCR' simulation is not an outlier -- but in fact typical for hydrodynamical simulations containing thermal feedback.

Fig.~\ref{fig:obs} shows that all the simulations without CRs produce \Lya spectra with significant flux at line center ($F_{\mathrm{valley}}\gtrsim 0.4 F_{\mathrm{peak}}$; see also the left panel of Fig.~\ref{fig:spectra}). This flux is reduced significantly from $t\gtrsim 30\,$Myr in the CR simulations as the outflow is established. Also the asymmetry of the line is increased leading to $L_{\mathrm{red}} / L_{\mathrm{blue}} \sim 2$. Both measures are more in line with what is found in observations where mostly $F_{\mathrm{valley}} / F_{\mathrm{peak}}\lesssim 0.1$ and $L_{\mathrm{red}} / L_{\mathrm{blue}} \gtrsim 2.5$ \citep[see also][]{Steidel2010ApJ...717..289S,Kulas2012ApJ...745...33K}.
Determining $F_{\mathrm{valley}}$ observationally requires a high spectral resolution since the convolution of the true spectrum with the instrument's kernel leads to a spreading of the peak flux, and thus, the true value of $F_{\mathrm{valley}}$ can be even lower than measured. This effect makes our comparison of $F_{\mathrm{valley}} / F_{\mathrm{peak}}$ conservative and the true discrepancy between the `noCR' simulation and the observations might be even larger.

In Fig.~\ref{fig:obs} we illustrate also how the emergent \Lya spectrum changes if the underlying  simulation had a lower resolution in the galactic halo which is typically the case for hydrodynamical simulations using adaptive refinement techniques or smoothed particle hydro-dynamics. In order to do this, we used the $t=70\,$Myr snapshot of the `noCR' simulation and averaged the hydrodynamical quantities in blocks of $500\,$pc sidelength one kiloparsec above and below the disk. The resulting spectral properties (pink circle in Fig.~\ref{fig:obs}) show that the values of $F_{\mathrm{valley}} / F_{\mathrm{peak}}$ and $L_{\mathrm{red}} / L_{\mathrm{blue}}$ are de- and increased compared to the unmodified `noCR' simulation, respectively. We discuss the implications that this purely numerical effect (smoothing due to low resolution) brings the simulated spectra in agreement with observations in Sec.~\ref{sec:discussion}.

While the inclusion of CR feedback lowers the tensions between the simulated spectra and the observed asymmetry as well as the flux at line center, this is only partially the case for the peak position of the red peak ($v_{\mathrm{red}}$ in the right panel of Fig.~\ref{fig:obs}). While CRs seem to increase this measure as well and, thus, bring it close to observed values, the latter can sometimes be as high as $v_{\mathrm{red}}\gtrsim 400\kms$ which we cannot reproduce (see the following discussion about possible origins of this discrepancy).

\begin{figure}
  \centering
  \includegraphics[width=.99\linewidth]{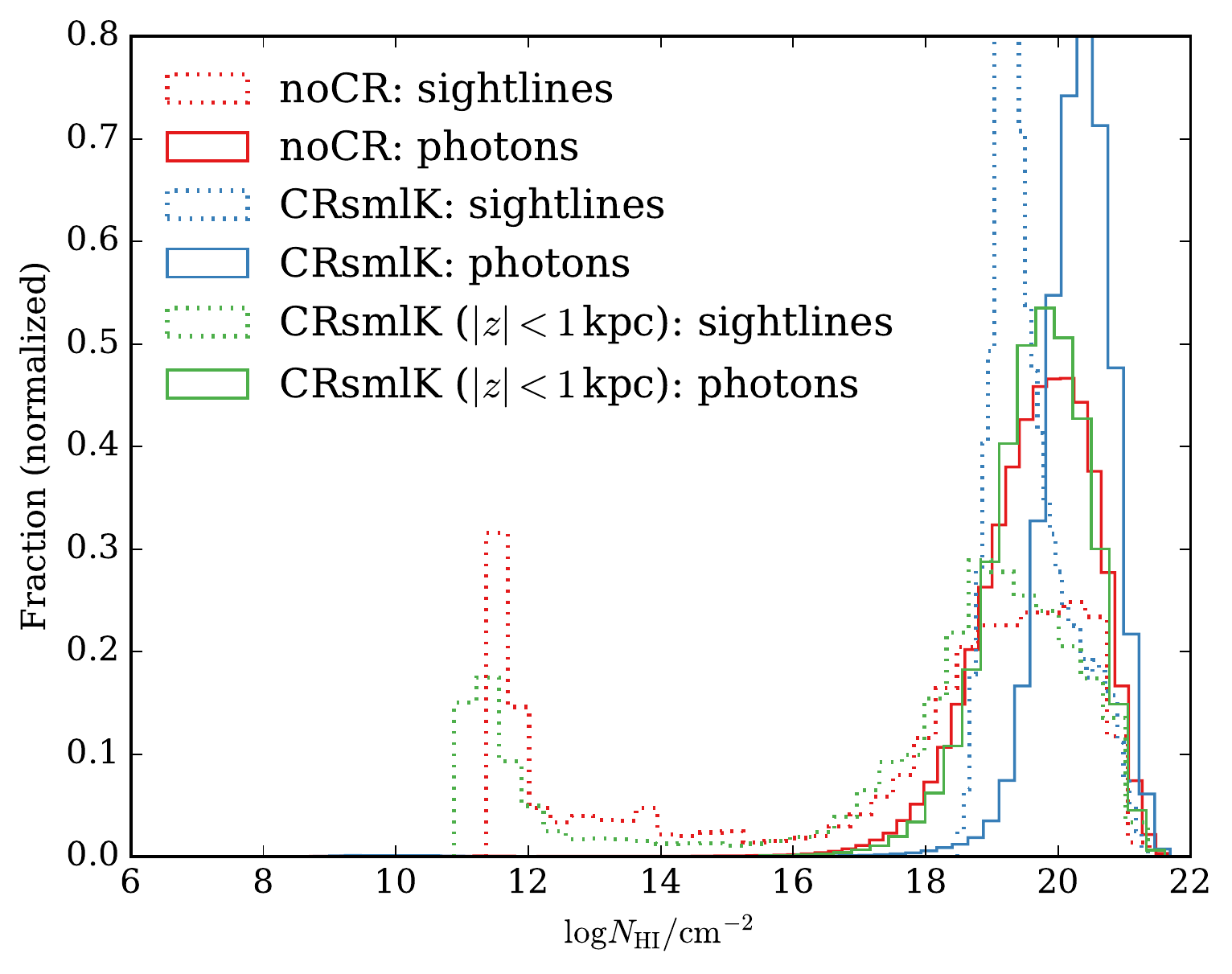}
  \caption{Distribution of neutral hydrogen column densities at $t=70\,$ Myr. The highly structured outflow (`noCR') results in a bimodal column density distribution. For illustration, we also added the distributions for the `CRsmlK' run but with the gas at $|z|>1\,$kpc removed which then resembles closely the `noCR' case.}
  \label{fig:dist}
\end{figure}

\section{Discussion \& conclusions}
\label{sec:discussion}

We have used the hydrodynamical simulations of the star-forming ISM also including magnetic fields and CRs as input for \Lya radiative transfer calculations and compared computed spectra to observations. Our results show that supernovae, stellar wind, and radiation feedback lead to spectra which are too symmetric and feature too much flux at line center compared to the observations \citep[also][]{Steidel2010ApJ...717..289S,Kulas2012ApJ...745...33K}. This is caused by low-density channels carved out by the SN feedback, which let \Lya photons escape at line-center, i.e., without first diffusing in frequency. CRs lead to a `shield' of neutral hydrogen outflowing at $\gtrsim 1\,$kpc height, which
prevents photons from escaping at line-center.

This is visible from Fig.~\ref{fig:dist} where we compare for a given time ($t = 70\,$Myr) the physical column density distribution (as dotted lines) to the column density crossed by the \Lya photons (solid lines in Fig.~\ref{fig:dist}). The column density crossed by the \Lya photons is larger than the physical one due to the scatterings and the resulting increase in pathlength (which leads to a widening of the intrinsic spectrum; cf. Fig.~\ref{fig:spectra}). Low-density channels cleared out by supernova feedback leave the physical $N_\HI$ distribution of the `noCR' simulation bimodal. 

Fig.~\ref{fig:dist} shows also that the photons scatter and diffuse in frequency. This can be seen from the example intrinsic spectrum in Fig.~\ref{fig:spectra} (left, dashed line), which is narrower than the emergent spectrum. Interestingly, this might explain the puzzling fact that the theoretical modelling of observed \Lya spectra require often a larger `intrinsic' line width than observed through H$\alpha$ \citep[e.g.][]{Yang2015,Orlitova2018arXiv180601027O}. Here, we show that through the turbulence of the inner disk, the line is widened prior to traversing the wind, yielding wider spectra than the intrinsic one even in the `noCR' case.\\

The difficulty in reproducing realistic \Lya spectra from galactic hydrodynamical simulations was noted earlier in the literature \citep[e.g.,][]{Gronke2017}. Several other causes for the discrepancy between the simulated \Lya spectra (without CR feedback) and the observations are possible:
\begin{itemize}
\item when comparing to observed \Lya spectra the partially neutral intergalactic medium might affect the observed \Lya line shape \citep{Dijkstra2007a}. In fact, the effect for $z\gtrsim 3$ is a decrease in flux at line center and a reduction of the blue peak -- just as ``required''. However, \textit{(i)} we compare the synthetic \Lya spectra also to $z\sim 0$ observations where the impact of the IGM on the \Lya line should be negligible \citep{Laursen2011}; and \textit{(ii)} even at higher redshift, an absorption of $\sim 50\%$ of the flux of \Lya emitting galaxies is unlikely because measured values of the \Lya escape fraction (or EWs) of individual galaxies \citep[e.g.,][]{Sobral2015} as well as the global average of all star-forming galaxies \citep{Hayes2011,2018MNRAS.476.4725S,2018PASJ...70S..14S} suggest higher transmission values.
  
\item the outer circumgalactic medium (CGM) can alter the \Lya line shape. To assess the importance of this effect a high-resolution ($\sim $pc) simulation covering a $\sim 10^3\,\mathrm{kpc}^{3}$ volume is required which would allow to compare modelled \Lya observables to data. \citet{Kakiichi2017} showed that observed \Lya halos surrounding star-forming galaxies \citep[e.g.,][]{Wisotzki2015} can be due to scattered \Lya radiation without altering the emergent inner \Lya spectrum heavily. However, more work in this direction is required.

\item small-scale, multiphase structure leads \Lya photons to escape as if the medium is homogeneous, resulting in spectra in agreement with observations \citep{Gronke2016b,Gronke2017}. An areal covering fraction of these small `droplets' of unity is required in order for this mechanism to work. This can be due to the in situ production of cold gas \citep[as, e.g., suggested by ][]{McCourt2016}, or the uniform transport of it without destruction -- which might be achieved through CRs demonstrating that the two proposed mechanisms are not exclusive.
\end{itemize}
Our setup has limitations \citep{Martizzi2016} but allowed us to highlight the potentially strong impact of CR driven winds on observed \Lya spectra. 
However, our idealized isolated setup ignores larger-scale effects such as filamentary inflows, and differential rotation \citep{2013ApJ...777L..38H,2016ApJ...824L..30P} on the properties of the cold outflow. Furthermore, the periodic boundary conditions might influence the results slightly. A study using a full galactic disk setup is required to overcome these limitations.
It is, however, crucial that this investigation is carried out with sufficient resolution also in the outer regions since -- as demonstrated in \S~\ref{sec:results} -- a too low resolution will remove the ionized channels and produce artificially ``correct'' spectra.

We show that the smooth-uniform outflows generated by CRs leave a clear imprint on the observable \Lya spectrum. For the flux at line center and the asymmetry of \Lya spectra, it is evident that the inclusion of CR feedback brings simulated \Lya spectra more in agreement with observations. Still, the agreement is not perfect: while our models with CRs produce positions of the red peak of $\sim 150\kms$, this value is lower than the bulk of observations which typically reach $v_{\mathrm{peak,red}}\sim 400\,\kms$. 
This might be due to the solar neighborhood conditions (gas surface densities of $\sim 10 M_\odot\,\mathrm{pc}^{-2}$), and correspondingly low SN rates of our hydro-dynamical simulations.

Nevertheless, our proposed picture of a cold, neutral CR-driven outflow surrounding the star-forming disk not only produces naturally the low flux at line center as well as the redward asymmetry of the line but also the widened `intrinsic' \Lya spectra sometimes required. All these points motivate further studies for the impact of CR feedback on \Lya observables.

\acknowledgments
MG thanks the organizers and participants of `SakuraCLAW', the MPA visitor program, and NASA grant NNX17AK58G. PG acknowledges funding from the ERC under ERC-CoG grant CRAGSMAN-646955.
TN and SW acknowledge support by the DFG priority program ``Physics of the interstellar medium''. SW greatly acknowledges funding by the ERC through the ERC Starting Grant no. 679852 RADFEEDBACK.
This research made use of \texttt{yt} \citep{Turk:2011}.

\bibliography{references_all}

\end{document}